\documentclass[twocolumn,secnumarabic,amssymb, nobibnotes, aps, prapplied,superscriptaddress]{revtex4-2}

\usepackage{graphicx}
\usepackage{siunitx}
\usepackage{amsmath}
\usepackage{mathtools}
\usepackage{xfrac}
\graphicspath{{Figures/}}
\usepackage{float}

\newcommand{\gtwo}{$g^{(2)}\left(\tau\right)$ }

\usepackage{enumitem}

\begin{document}

\title{Measuring photon correlation using imperfect detectors}

\author{Rachel N. Clark}
\affiliation{School of Engineering, Cardiff University, Queen's Building, The Parade, Cardiff, CF24 3AA, UK}
\affiliation{Translational Research Hub, Maindy Road, Cardiff, CF24 4HQ, UK}

\author{Sam G. Bishop}
\affiliation{School of Engineering, Cardiff University, Queen's Building, The Parade, Cardiff, CF24 3AA, UK}
\affiliation{Translational Research Hub, Maindy Road, Cardiff, CF24 4HQ, UK}

\author{Joseph K. Cannon}
\affiliation{School of Engineering, Cardiff University, Queen's Building, The Parade, Cardiff, CF24 3AA, UK}
\affiliation{Translational Research Hub, Maindy Road, Cardiff, CF24 4HQ, UK}

\author{John P. Hadden}
\affiliation{School of Engineering, Cardiff University, Queen's Building, The Parade, Cardiff, CF24 3AA, UK}
\affiliation{Translational Research Hub, Maindy Road, Cardiff, CF24 4HQ, UK}

\author{Philip R. Dolan}
\affiliation{National Physical Laboratory, Hampton Road, Teddington, TW11 0LW, UK.}

\author{Alastair G. Sinclair}
\affiliation{National Physical Laboratory, Hampton Road, Teddington, TW11 0LW, UK.}

\author{Anthony J. Bennett}
\email{BennettA19@cardiff.ac.uk}
\affiliation{School of Engineering, Cardiff University, Queen's Building, The Parade, Cardiff, CF24 3AA, UK}
\affiliation{Translational Research Hub, Maindy Road, Cardiff, CF24 4HQ, UK}

\date{\today}%
\begin{abstract} 

Single-photon detectors are ``blind" after the detection of a photon, and thereafter display a characteristic recovery in efficiency, during which the number of undetected photons depends on the statistics of the incident light. We show how the efficiency-recovery, photon statistics and intensity have an interdependent relationship which suppresses a detector's ability to count photons and measure correlations. We also demonstrate this effect with an experiment using $n$ such detectors to determine the $n^{\mathrm{th}}$ order correlation function with pseudothermal light.

\end{abstract}

\maketitle

\section{Introduction}
\label{sec:intro}
Many photonic quantum technologies depend on efficient detectors which are sensitive to the arrival of a single quanta of light \cite{hadfield,chunilall}. Photons are an ideal qubit, due to their many degrees of freedom, low decoherence and high speed \cite{divincenzo,bouwmeester}. In recent decades single-photon detectors have driven fundamental tests of quantum mechanics, from the original Bell tests \cite{belltest3}, through to tests of non-locality \cite{nonlocality} and local realism \cite{localrealism}. In addition, the continued development of these detectors is driving advances in range-finding \cite{Gariepy2015}, advanced imaging \cite{Genovese_2016}, quantum-secure communications \cite{gisin2002quantum}, and photonic quantum computing \cite{knill}. At the heart of these technologies, the concepts of measurement, correlation and metrology are essential to verify and implement the operations. Here we study how real single-photon detectors count photons and quantify the fidelity with which they measure intensity and correlation.   

Single-photon detectors, such as Avalanche Photo Diodes (APDs) based on silicon \cite{ruegg} or InGaAs \cite{InGaAsSPAD}, can be purchased in convenient Peltier-cooled portable units to measure photon arrival times within a few hundred picoseconds in Geiger mode. After each photon detection, these devices must be reset by sweeping away carriers generated in the avalanche, which is usually achieved with an embedded electrical circuit. During this reset, which typically takes tens of nanoseconds, the device is unable to count photons, preventing measurement of a light source's statistics with a single detector. This is known as the detector's `dead time'. Immediately following the dead time, there is an associated `reset time' before the detection efficiency fully recovers \cite{chunilall}. Hereafter, we call the full variation in efficiency after a detection event the temporal efficiency-recovery (TER). More recently, superconducting nanowire single-photon detectors (SNSPDs) \cite{pico} have been developed that offer efficiencies above 90\% \cite{reddy2020superconducting}, timing accuracy of tens of picoseconds, and sub-Hz dark count rates. 
These devices consist of a meandering superconducting wire\cite{design} biased below its critical current. Upon absorption of a single photon, a resistive barrier forms and the bias current is shunted across a resistor. The kinetic inductance of the nanowire leads to a TER lasting tens of nanoseconds, during which the supercurrent recovers to its normal efficiency \cite{hadfield}.

It is an on-going challenge to develop single-photon and multi-photon light sources with high brightness. Semiconductor quantum dot devices have been shown to give rise to detection rates approaching \SI{50}{\mega\hertz} \cite{2021bright}. Both quantum dots and spontaneous parametric down conversion sources can now be used to execute certain algorithms at speeds which challenge classical computers  \cite{tillman, doi:10.1126/science.abe8770, wang2017high}. At these high photon rates commercial single-photon detectors saturate, impeding their ability to count all detection events \cite{Ann2015, Beck2007}.

In this work, we theoretically consider the TER-induced saturation of single-photon detectors at high rates with different photon statistics. We simulate the response of these detectors using the concept of the waiting time distribution, $ \Omega\left(\mathrm{dt}\right) $, which describes the distribution of time intervals, $ \left(\mathrm{dt}\right) $, between consecutive detections. Using $ \Omega\left(\mathrm{dt}\right) $ we characterize the TER of a SNSPD showing it is not an abrupt step, but a smooth time-varying function \cite{chunilall}. Furthermore, we clarify how the TER and source statistics must both be considered to quantify the efficiency. Finally, we show how TER impacts measurement of positive multi-photon correlation, such as a generalized version of the famous Hanbury-Brown and Twiss experiment \cite{HBT} to determine the $n$\textsuperscript{th} order photon correlation function. In this experiment a pseudothermal light source is used, whose intensity can be changed whilst retaining the same photon statistics. These results show that detector imperfections must be factored into the analysis of photon correlation, whether it be in quantifying the degree of entanglement, in quantum light source metrology, or in optical read out of quantum photonic technologies, which will be of increasing importance as sources and detectors are developed to operate at higher rates.

\section{Temporal efficiency recovery of detectors}
\label{sec:Model}
\begin{figure}[h]
    \includegraphics[width=8.6cm]{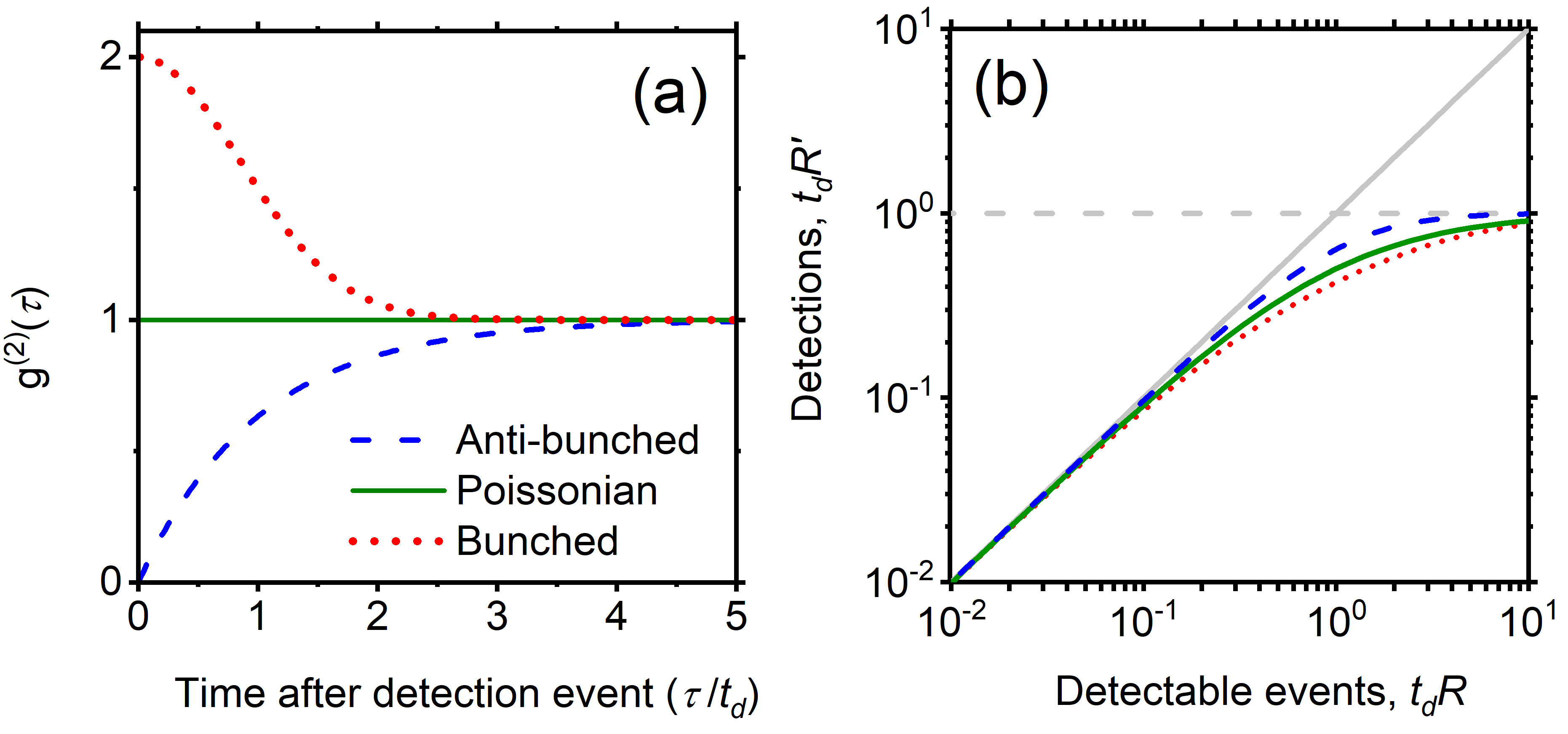} 
        \caption{Photon statistics of a light source determines how the detected photon rate varies with increasing incident flux. (a) Probability of a second photon being incident on the detector at time $\tau$ after an initial detection event at time $\tau=0$ relative to a Poissonian source (green, solid), shown for bunched thermal (red, dotted) and 2-level antibunched source (blue, dashed) with characteristic timescales equal to the detector response time $t_{d}$. (b) Simulation of unitless detected photon rates ($t_{d}R'$) for the three sources, as a function of the incident flux ($t_{d}R$). A straight grey line shows the detection rate in the absence of TER. Horizontal dashed line shows the saturated rate.}
	\label{fig:Fig1}
\end{figure}

Consider a single-photon detector with a TER described by a Heaviside function after a detection event, during which the detector is completely ``blind" and then instantaneously recovers at time $\tau = t_{d}$, where $t_{d}$ is the dead time. Therefore, all photons arriving after an initial detection at time zero in Fig. \ref{fig:Fig1}(a) up until $t_{d}$, are not registered. It follows that, regardless of source statistics, a detector with this behaviour must have a saturated detection rate no greater than $t_{d}^{-1}$. 

Fig. \ref{fig:Fig1}(a) also shows the relative probability of a second photon being incident on the detector at time $\tau$ after an initial detection event at time $\tau=0$, for two archetypal light sources (in blue and red), compared to a Poissonian source of the same intensity (in green). This relative probability is equal to the second-order correlation function, \gtwo, measured in the Hanbury-Brown and Twiss experiment from the distribution of photon arrivals at two independent detectors \cite{HBT}. Its form classifies three different signatures of photon statistics: Poissonian, bunched and antibunched. In their simplest form these are, respectively: 
\begin{align}
    g^{\left(2\right)}\left(\tau\right) &=  1, \\
    g^{\left(2\right)}\left(\tau\right) &=  1 + \exp\left[ \frac{- \mathrm{ln}\vert 2\vert \cdot \tau^{2}}{T^{2}}\right], \label{eq:g2Bunched}\\
    g^{\left(2\right)}\left(\tau\right) &=  1 - \exp\left[\sfrac{-\vert \tau \vert}{T} \right], \label{eq:g2ABunched}
\end{align}
where $T$ is the characteristic timescale of the bunching/antibunching, and the form of the exponent is chosen such that the full width half maximum of functions described by Equations \ref{eq:g2Bunched} and \ref{eq:g2ABunched} are equal. To illustrate this, in Fig. \ref{fig:Fig1}(a) we set $T=t_{d}$. Since the temporal distribution of photons from each source differs, it follows that the number of photons `missed' during the detector recovery also varies.  For a Poissonian light source the detection rate reduces to the simple form
\begin{equation}
    R' = R/\left(1 + R\cdot t_{d}\right),
    \label{eq:Rdash}
\end{equation}
where $R$ is the detectable event rate in the absence of TER. For a bunched light source each detection event initiates a TER during which the detector misses a larger fraction of the incident photons, and the detection rate $R'$ is therefore lower than the Poissonian case, even when the source intensities are the same. Conversely, for an antibunched source a smaller fraction of photons are missed during the TER, so the detection rate $R'$ is higher than the Poissonian case.  A calculation of the variation in detected photon rate for the three cases is shown in Fig. \ref{fig:Fig1}(b) with the rates presented in unitless form (normalized to $t_{d}^{-1}$). For comparison a dashed horizontal line at the saturation detection rate is included. Details of this calculation can be found in the Supplementary Material. We omit the effect of dark counts on the detector behaviour, as the dark rate is below 1Hz in our experiment. Previously, authors have studied the effects of dark counts on detector performance, which are important at low photon fluxes \cite{georgieva2021detection}. 

\begin{figure*}[ht!]
    \includegraphics[width=0.9\linewidth]{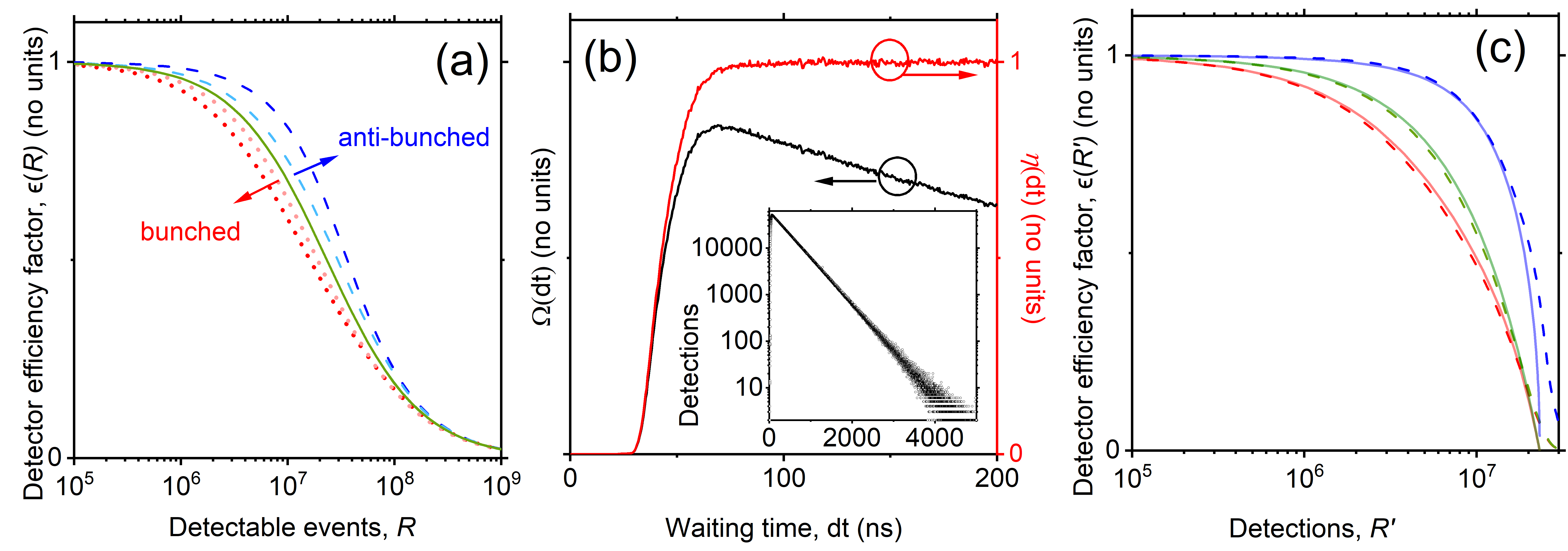}
    \caption{Understanding the rate-dependent efficiency of a real single-photon detector. (a) Detector efficiency factor $\epsilon(R)$ as a function of the detectable event rate ($R$) in the absence of TER. Poissonian light source source (green, solid line). Bunched thermal light source with temporal HWHM $t_{d}$ (red, dotted) and $t_{d} /4$ (pink, dotted).  Antibunched light sources with  timescale $T=t_{d}$ (blue, dashed), $t_{d}/4$ (pale blue, dashed).(b) Experimental waiting time distribution (black) and temporal efficiency recovery (TER), $\eta\left(\mathrm{dt}\right)$ (red). Insert shows the raw data used to determine the waiting time distribution. (c) Detector efficiency factors for detectors with abrupt (solid lines) and experimental (dashed) TER response, for Poissonian (green), Bunched (red) and antibunched (blue) photon statistics, as a function of the detection rate, $R'$.}
    \label{fig:Fig2}
\end{figure*}

The suppressed ability of a typical single-photon detector to register further photons within its TER leads to a rate-dependent efficiency factor, in  addition to its intrinsic internal quantum efficiency. We define an efficiency factor, $ \epsilon(R) = R'/ R$, which is the fraction of photons detected relative to the photon rate from an otherwise identical detector without TER. Fig. \ref{fig:Fig2}(a) shows simulations of $\epsilon\left(R\right)$ (see Supplementary Material for more details of the model) for the case when $t_{d}=$ \SI{43}{\nano\second} which we show later is typical for detectors used in our experiments. The green solid line shows the case for a Poissonian state extracted from the model
\begin{equation}
    \epsilon\left(R\right) = \left(1 + R\cdot t_{d}\right)^{-1},
    \label{eq:EfficiencyFactor}
\end{equation}
which agrees with the expected form as in Equation \ref{eq:Rdash}, and reaches a value of 0.5 at $R = t_{d}^{-1}$, for the Poissonian source. A red dotted line shows $\epsilon\left(R\right)$ for the thermal source shown with timescale $t_{d}$, and the pink dotted line shows predictions for a thermal source with a HWHM $t_{d} / 4$. Conversely, blue and pale-blue dashed lines show the results for antibunched sources with timescale $t_{d}$ and $t_{d} / 4$, respectively. We see that $\epsilon\left(R\right)$ converges on the Poissonian case for sources with bunching and antibunching at faster timescales. Several sources do have timescales relevant to the $\sim 40$\si{\nano\second} TER of these commercial SNSPDs: semiconductor quantum dots display antibunching on the nanosecond timescale under non-resonant excitation \cite{2021bright}, but have been reported to display complex bunching behaviour on nanosecond to millisecond times due to charge trapping \cite{Davanco2014}. By contrast, emission from the nitrogen vacancy center in diamond displays antibunching and bunching on timescales of several tens of nanoseconds \cite{Kurtsiefer2000}.

\begin{figure*}[ht!]
    \includegraphics[width=0.8\linewidth]{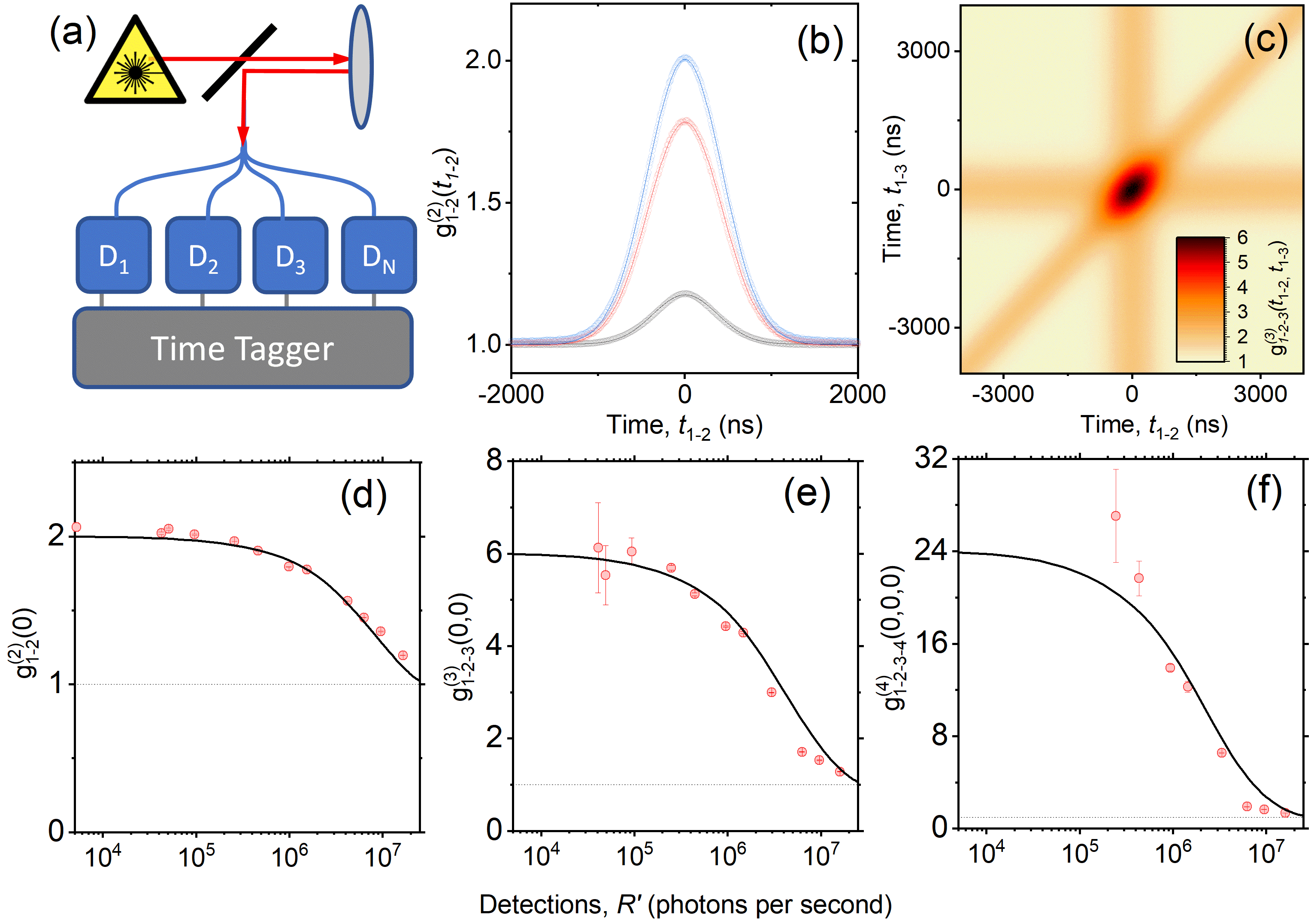}
    \caption{The effect of TER on the higher order photon correlations of a pseudothermal source. (a) Apparatus used to measure the higher order correlations. (b) Second order correlation function at detected photon rates of $\sim 10^{5}$ cps (blue circles), $\sim 10^{6}$ cps (red circles) and $\sim 10^{7}$ cps (black circles). Each dataset is fitted with a Gaussian  of width $\sim$\SI{418}{\nano\second} and varying amplitude (solid lines) (c) third-order correlation function at a detected photon rate of $\sim$ \num{1d4} cps. The lower row of panels shows the maxima of the (d) second order correlation function, (e) third order correlation function and (f) fourth order correlation function, for coincident detections as a function of detected photon rate, $R'$. Results of simulations shown as a solid line. Errors are estimated from the square-root of the number of coincidence events.}
    \label{fig:Fig3}
\end{figure*}

We determine the TER of our commercial SNSPDs, $ \eta\left(\mathrm{dt}\right)$, from measurements that time-tag every photon detection event from a Poissonian source attenuated to a rate well below $t_{d}^{-1}$, as follows. From the time-tagged data file we calculate the time difference, $\mathrm{dt}$, between consecutive detection events only and subsequently the distribution of these consecutive time differences, which is the waiting time distribution, $\Omega\left(\mathrm{dt}\right) $. Fig. \ref{fig:Fig2}(b) shows that at times much greater than $t_{d}$ the waiting time distribution  $\Omega\left(\mathrm{dt}\right)$  has an exponentially decaying trend with exponent determined by the photon detection rate (black line). Normalising out this exponential trend reveals the response of the detector immediately after a photon detection, $\eta(\mathrm{dt})$ in Fig. \ref{fig:Fig2}(b) (red line). For the SNSPDs under test (IDQuantique ID281) the TER has a smoothly varying form that returns to 50\% of its long term value at \SI{43}{\nano\second}. SNSPDs often display a bias dependent TER, which can distort the temporal response of the detectors immediately after a preceding pulse \cite{Craiciu} as supercurrent returns to the detector. For the purposes of this work we assume the TER is constant, which we show provides an adequate fit to experimental data in Fig. \ref{fig:Fig3}.

The TER in Fig. \ref{fig:Fig2}(b), $\eta\left(\mathrm{dt}\right)$, leads to a small deviation in the saturation curve, relative to the predicted response for a detector with a Heaviside TER. To simulate this we use a numerical model of the waiting time distribution, using the experimentally determined $ \eta\left(\mathrm{dt}\right)$ (see Supplementary Material). The inverse of the mean-waiting time, $\langle \Omega\left(\mathrm{dt}\right)\rangle^{-1}$ is the detection rate, $R'$. Fig. \ref{fig:Fig2}(c) shows $\epsilon(R')$ for the three sources described in Fig. \ref{fig:Fig1}(a), when the detector has a Heaviside TER (solid lines) and with the real detector response (dashed lines). We see the data has similar $\epsilon(R')$ for the two TER responses below $t_{d}^{-1}$. However, the SNSPD has some finite efficiency at times below $t_{d}$, which ensures that it is possible, albeit at low efficiency, to register detection events at rates above $t_{d}^{-1}$. It is at these highest detection rates the true TER response has the greatest impact on the photon detection rate and correlation. However, as we will show, the TER can distort the correlation two orders of magnitude below the saturated rate.

\section{Intensity-dependent photon correlation}
We now show experimentally that the degree of correlation is intensity dependent as a result of this TER-induced saturation. Pseudothermal light was generated from the speckle of light reflected from a rotating ground glass diffuser and split equally between an array of comparable SNSPDs to measure the second, third and fourth-order correlation functions, Fig. \ref{fig:Fig3}(a). Examples of \gtwo are shown in Fig. \ref{fig:Fig3}(b) at detected rates of $\sim 10^{5}$  counts-per-second (cps) where $g^{(2)}(0) = 2.00$ (blue), $\sim 10^{6}$ cps where $g^{(2)}(0) = 1.79$ (red) and $\sim 10^{7}$ cps where $g^{(2)}(0) = 1.18$ (black) - showing a reduced bunching amplitude at higher photon rates. In these measurements the photon statistics have not changed, only the incident photon rate, and yet the correlation is suppressed at rates more than two orders of magnitude below saturation. 

In some applications it is insufficient to measure the second-order correlation and higher-order correlations are required to understand the internal dynamics of a source \cite{Rundquist14} or to implement a quantum protocol, such as teleportation \cite{bouwmeesterTel}. Using the same apparatus we measured a series of higher-order correlations from the pseudothermal source, such as the third-order correlation shown in Fig. \ref{fig:Fig3}(c). At photon rates below $10^{5}$ cps (more than two orders of magnitude below the detectors' saturated rates) the amplitude of the correlation functions $g^{(n)}(0,0...)$ approach the expected values $n!$ for a pseudothermal source. However, at higher detection rates we observe the amplitude of the second-, third- and fourth-order correlations $g^{(n)}(0,0...)$ are strongly dependent on detection rate, as shown in Fig. \ref{fig:Fig3}(d-f), tending to unity at saturation of the detector.

We now prove these observations are a direct result of the rate-dependent efficiency of single-photon detectors. The measurements in Fig. \ref{fig:Fig3} arise from correlations between the detected photon rates, $R'$, so to predict $g^{(n)}_{\mathrm{exp}}(0, 0...)$ we can use a generalized $n$th-order function:

\begin{equation} 
g^{(n)}_{\mathrm{exp}}(0,0...) = \dfrac{\langle  \prod^{n}_{i=1}R'_{i} \rangle}{ \prod^{n}_{i=1} \langle R'_{i} \rangle } = \dfrac{\langle  \prod^{n}_{i=1}\epsilon_{i}(R_{i})\cdot R_{i} \rangle}{ \prod^{n}_{i=1} \langle \epsilon_{i}(R_{i})\cdot R_{i} \rangle },
\label{eq:gn}
\end{equation}
where triangular brackets denote the mean and $\epsilon_{i}(R_{i})$ is the detection efficiency factor for the $i$\textsuperscript{th} detector. It is well known that the intensity (and thus rate, $R$) distribution of thermal light, $\xi\left(R\right)$, is:
\begin{equation} 
\xi\left( R \right) \propto \frac{1}{\langle R \rangle}\exp\left[\frac{-R}{\langle R\rangle}\right],
\label{eq:eq2}
\end{equation} 
which can be used to determine the mean values in Equation \ref{eq:gn}. Further information on the evaluation of Equation \ref{eq:gn} is given in the Supplementary Material. We assume the light is split equally between identical detectors to generate the predicted curves in Fig. \ref{fig:Fig3} (d)-(f) showing $g^{(n)}(0, 0...)$ as a function of the photon detection rate, $R'$. The experimental data follows the predicted correlation functions which tend to $n!$ at low $R'$ and to unity at high $R'$. Deviations between the experimental data and the predicted curve may be a result of using periodic speckle to generate pseudothermal light that does not perfectly follow Equation \ref{eq:eq2} (see Supplementary Material for further discussion).  

We determine $\eta\left(\mathrm{dt}\right)$ by averaging over many events at $R' << t_d^{-1}$, and assume the smoothly varying TER is identical for each event and $R'$. However, the nanowire detector has an inhomogeneous width, which would lead to an absorption and position dependent response, on average reproducing $\eta\left(\mathrm{dt}\right)$. We leave experimental investigation of this effect to future work. However, our experiment shows that the time-averaged TER is key to understanding the rate-dependence of $g^{(n)}(0, 0...)$.

\section{Conclusion and outlook}
\label{sec:conc}
In the future, measurement systems can be designed to ameliorate this issue by ensuring light incident on each detector displays Poissonian statistics, even if the photons impinging on separate detectors are correlated, such as by using non-degenerate entangled photon pair sources and separating the signal and idler. Additionally, the development of photon detectors with faster TER, using active electronics to quench the SNSPD \cite{Ravindran:20}, will reduce the magnitude of the effect. Another approach is the development of arrayed multipixel single photon detectors \cite{perrenoud2021operation}. If we consider the measurement of \textit{n}-fold correlation using a \textit{m}-fold splitter and \textit{m}-detectors, we can sum all \textit{n}-fold correlations within the array. 
\begin{figure}[h]
    \includegraphics[width=8.6cm]{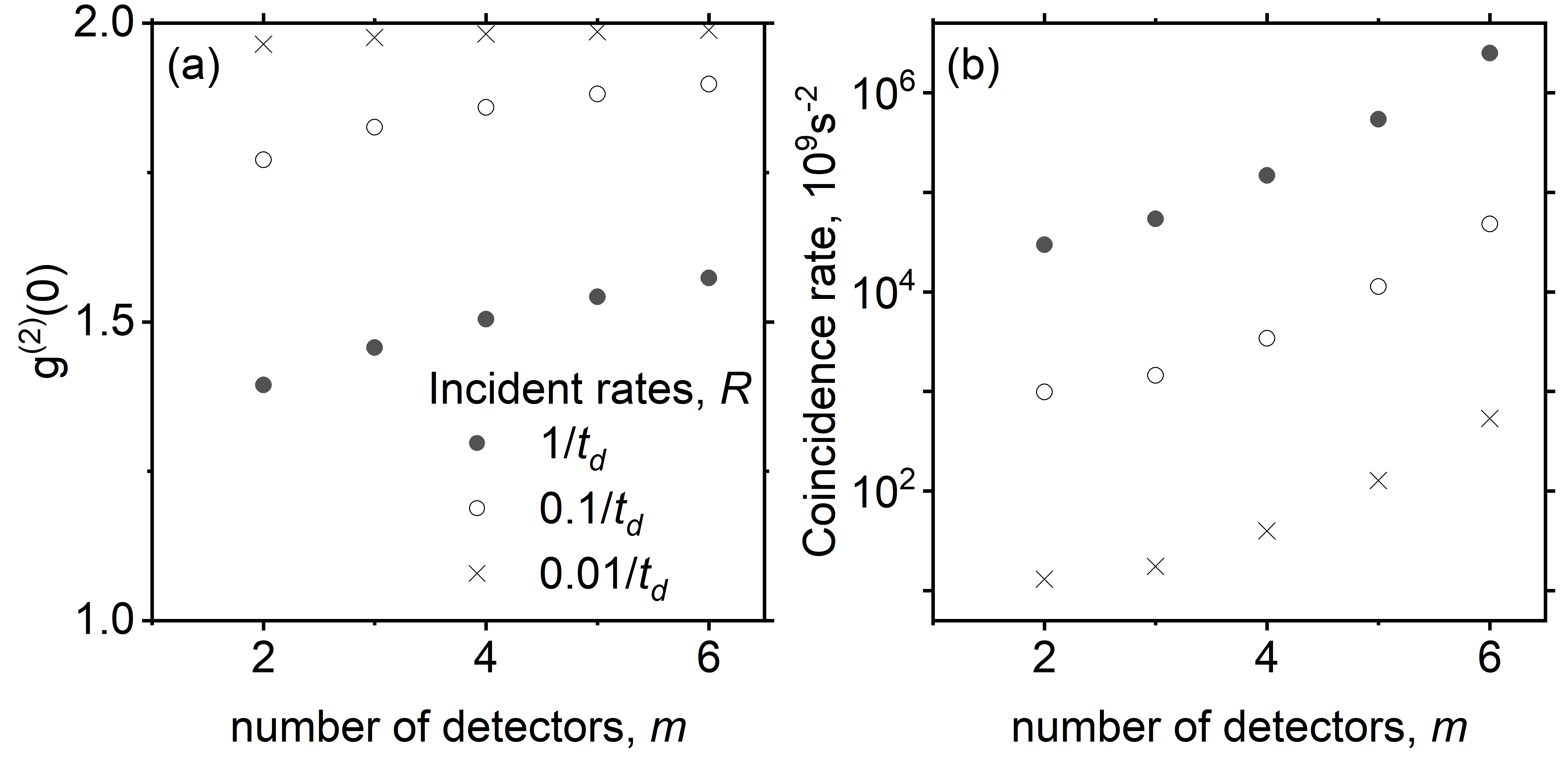} 
    	\caption{Measurement of second-order correlation with an array of \textit{m}-detectors. (a) $g^{(2)}(0)$ predicted from summation of pair-wise correlations across \textit{m}-detectors using a \textit{m}-way splitter, for 3 different incident photon rates. (b) Corresponding rate at which the correlation accumulates per nanosecond time bin.  }
	\label{fig:Fig4}
\end{figure}
The coincidence rate, $C(n)$, scales as: 
\begin{equation} 
C(n) \propto  \dfrac{ m!}{n!} . \left[ \dfrac{\epsilon(R/m)\cdot R}{m} \right]^{n},
\label{eq:Cn}
\end{equation}
for high photon rates and low numbers of detectors. Fig. \ref{fig:Fig4} shows a simulation of the $g^{(2)}(0)$ value of thermal light and $C(n=2)$ as a function of \textit{m}, the number of ways the light is split, for three different fluxes near saturation. Increasing \textit{m} reduces the photon rate on each detector, improving the fidelity of the correlation to the real value. The addition of more detectors reduces the coincidence rate for a given detector-pair, but the summation across all pairs gives an advantage in acquisition rate at high \textit{m}, albeit at a considerable cost to hardware complexity. This should act as a strong motivation for the development of multi-pixel single-photon detectors.

In conclusion, our analysis is essential for the development of high photon rate quantum technologies and will have significant consequences when the degree of correlation is used in protocols for secure communications \cite{ekert} or for tests of quantum fundamentals \cite{PhysRevLett.23.880}. We have shown that photon rates and correlations are underestimated by detectors operating within two orders of magnitude of their saturated rate. Thus, to accurately measure a source with a detector close to its saturation rate, it is essential to carefully calibrate (i) the TER of the detector, which is not simply parameterized by a single $t_d$ value, (ii) the detector's internal quantum efficiency long after a detection event and (iii) the statistics of the source. As quantum photonic technologies employing high photon rates continue to be developed, it is important to properly account for imperfections in detector performance.

\section*{Acknowledgements}
We acknowledge financial support provided by EPSRC via Grant No. EP/T017813/1 and EP/T001062/1. RC was supported by grant EP/S024441/1, Cardiff University and the National Physical Laboratory. We thank Dr Jonathan Fletcher of the National Physical Laboratory for an internal review.

\appendix 
\section{Experimental methods}
\label{sec:AppendixMethods}
\begin{figure}
    \includegraphics[width=0.95\linewidth]{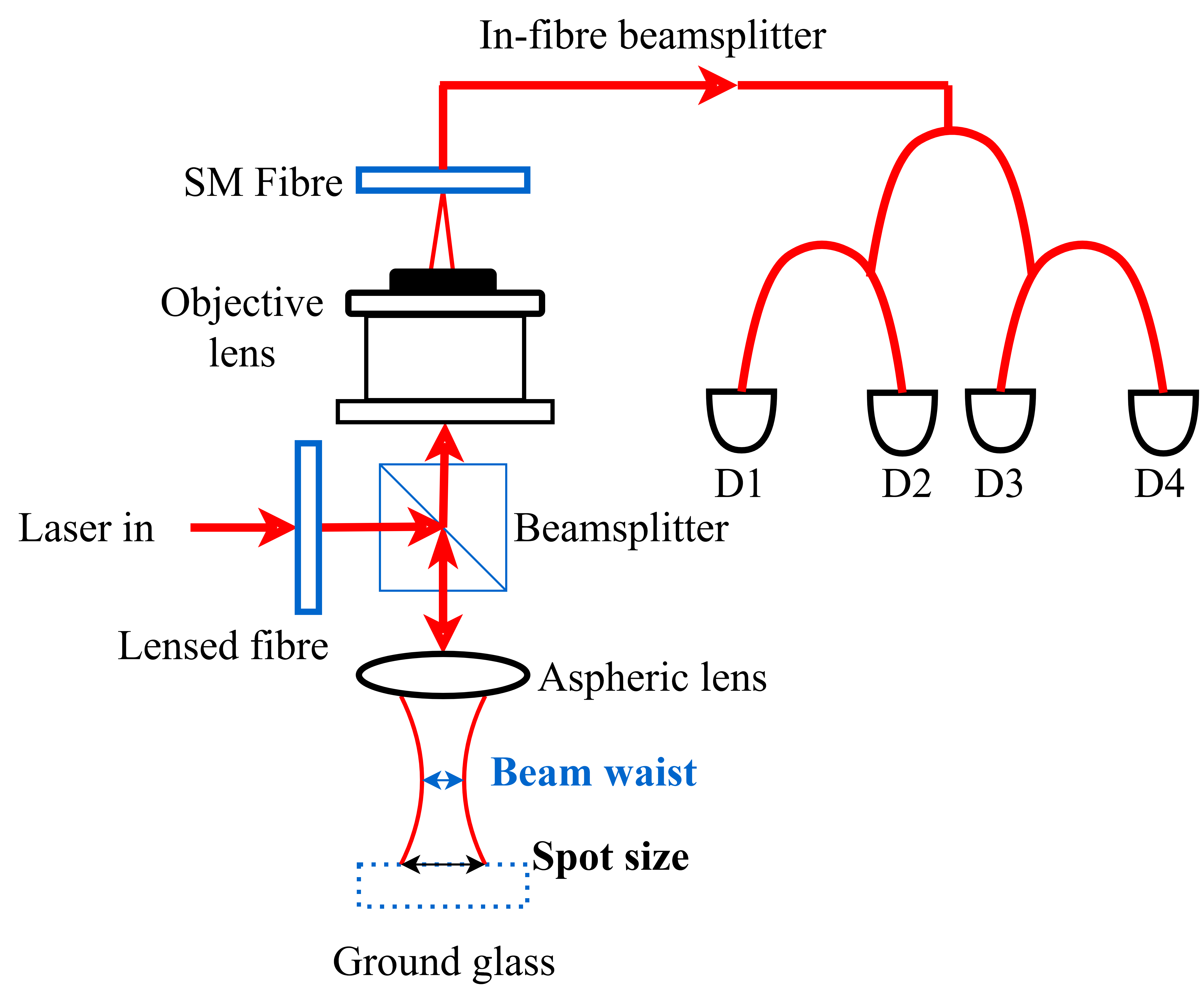}
    \caption{A schematic of the pseudothermal light experiment. Coherent light from a PicoQuant \SI{850}{\nano\metre} laser is expanded with a lensed fibre and then focussed onto a rotating ground glass, where the focussing lens has been carefully selected to ensure Gaussian statistics are recovered. The reflected pseudothermal light is collected into the objective and shared between four SNSPD detectors via in-fibre directional couplers.}
    \label{fig:pseudothermal}
\end{figure}

The experimental system for generating pseudothermal light with bunched statistics is shown in Figure \ref{fig:pseudothermal}. A PicoQuant laser diode at \SI{850}{\nano\metre} operating in CW mode is focussed onto to a rotating ground glass diffuser of 1500 grit, via an aspheric lens of focal length \SI{18.4}{\milli\metre} leading to a spot size of $\sim$\SI{10}{\micro\metre}. After being reflected and scattered by the rotating ground glass, the light was collected into a second fiber via a Mitutoyo objective (NA $= 0.2$), and then directed into an array of  three $2\times2$ directional couplers, used to split the light between four detection channels (D1-4). The detectors were ID281 SNSPDs in a closed cycle cryostat supplied by ID Quantique, which have an efficiency of $\geq 90\%$ at this wavelength, and an intrinsic timing accuracy of $\sim$\SI{30}{\pico\second}. Time-tagged photon streams were recorded with \SI{100}{\pico\second} binning using ID900 time-correlated photon counting electronics. Subsequently, the time-tagged data was analysed in MATLAB using custom software.

\begin{figure}
    \centering
    \includegraphics[width=\linewidth]{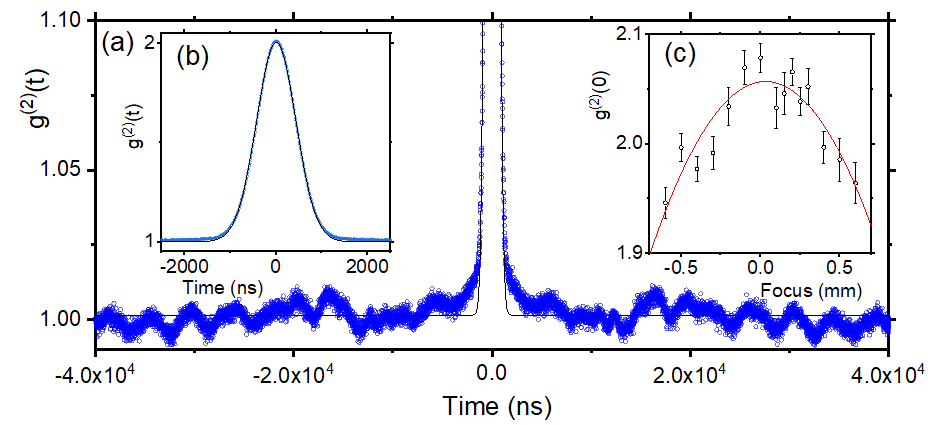}
    \caption{Imperfections in the pseudothermal light source. (a) shows a zoom of the correlations at long times (b) shows the approximately gaussian peak within a few microseconds of coincidence and (c) the amplitude of the correlation as a function of the position of the rotating ground glass.}
    \label{fig:S2}
\end{figure}

The apparatus of Fig. \ref{fig:pseudothermal} generates light with some characteristics of thermal light, such as an approximately Gaussian second-order correlation function near time zero with maximum value close to two, as shown in Fig. \ref{fig:S2}(b). The incoming light is scattered by the grit of the rough ground glass; for a large number of randomly distributed  scattering centres interference results in an intensity variation in the far field known as \textit{speckle}\cite{oliver1963sparkling,sporton1969scattering}. The Gaussian nature of the intensity fluctuations is only observed when the spot illuminates a sufficiently high number of scatterers - if the spot is focused too tightly, the intensity fluctuations become abrupt and the complex speckle pattern is not produced \cite{bluemel1972photon}. Thermal light produced in this manner is monochromatic and has a tunable coherence time. Jakeman \cite{jakeman1975effect} and Pusey \cite{pusey1976photon} investigated the physics of the speckle pattern and demonstrated that its coherence time is related to the in-plane velocity of the spot on the ground glass and the spot size of the incident laser light. Kuusela \cite{kuusela2017measurement} performed an in-depth study of the relation between different optical parameters and the statistical nature of the generated pseudothermal light, for example, the distance of the beam waist from the plane of the ground glass, and the intensity variations of the output light for different ground glass grit.

Inspection of the correlation function at long times reveals structure within 0.01 of unity, shown in Fig. \ref{fig:S2}(a) which is indicative of periodic fluctuations in the speckle pattern as the ground glass rotates. A further imperfection arises from the focusing of the light onto the ground glass, which can lead to an amplitude which is not exactly two (Fig. \ref{fig:S2} (c)). In the main text we position the rotating ground glass such that the spot is slightly defocussed, producing $g^{(2)} (0) = 2$ \cite{kuusela2017measurement}. 

\section{Numerical simulation of waiting time distribution}
\label{sec:AppendixModel}
The waiting time distribution, $\Omega(\mathrm{dt})$, is numerically simulated using: 
\begin{equation} \label{eqomega}
\Omega (\mathrm{dt} + \delta(\mathrm{dt}))  = \Omega(\mathrm{dt}) - \Omega (\mathrm{dt})\cdot\eta(\mathrm{dt})\cdot P(\mathrm{dt})\cdot\delta(\mathrm{dt}),
\end{equation}
where $\mathrm{dt}$ is the time since the initial detection, $\delta(\mathrm{dt})$ is a infinitesimal time interval, $\eta(\mathrm{dt})$ the detector TER function and $P(\mathrm{dt})$ the probability of a second photon arriving at the detector. In future, an analytical fit to $\eta(\mathrm{dt}$ derived from first principles could be used to derive a closed form for the waiting time distribution.

Additionally, the mean waiting time is given by: 
\begin{equation}\label{eqdt} 
\overline{\mathrm{dt}} = \dfrac{\int_0^{\infty} \mathrm{dt}\cdot\Omega(\mathrm{dt}).\delta(\mathrm{dt})}{\int_0^{\infty}  \Omega(\mathrm{dt}).\delta(\mathrm{dt}) }.
\end{equation}

Consider the case of constant $\eta(\mathrm{dt}) = \eta_{0}$, for the case of a detector with no TER. For the Poissonian light source $P(\mathrm{dt})= P_{0}$, we can use a trial function $\Omega(\mathrm{dt}) = e^{-R\cdot \mathrm{dt}}$, which reveals $R = \eta_{0}.P_{0}$. In other words, the detection rate in the absence of TER for a Poissonian source is the exponential decay rate of $\Omega(\mathrm{dt})$. It can then be shown that $\overline{\mathrm{dt}} ^{-1} = R$, as expected.

In this work we consider the combination of non-constant TER, $\eta(\mathrm{dt})$, and non-constant $P(\mathrm{dt})$ as a result of the source's photon statistics. We assume the probability of their being a second incident photon $P(\mathrm{dt})$  after the initial detection event is proportional to the product $ g^{(2)}(\mathrm{dt})\cdot I$ where $I$ is the incident flux. To create the curves in Fig. \ref{fig:Fig1}(b), \ref{fig:Fig2}(a) and \ref{fig:Fig2}(c) we perform two numerical simulations for each $g^{(2)}(\mathrm{dt})$, at a range of values of $I$, to determine $\Omega(\mathrm{dt})$:  

\begin{enumerate}[label =\roman*.]
    \item with detectors having some time-varying TER, $\eta(\mathrm{dt})$. From $\Omega(\mathrm{dt})$ we determine the mean waiting time, and $R'$ as a function of $I$.
    \item with otherwise identical detectors but which do not have any TER. From this $\Omega(\mathrm{dt})$ we determine the mean waiting time, and $R$ as a function of $I$.
\end{enumerate}

Fig. \ref{fig:Fig1}(b) plots the results of these two simulations of $R'$ against $R$. Additionally, the time-averaged detector efficiency function $R'/R$ is shown as a function of $R$ in Fig. \ref{fig:Fig2}(a). 

In the case of the experimentally determined detector TER shown in Fig. \ref{fig:Fig2}(b), the numerical simulation proceeds in a similar manner to produce the dashed curves in Fig. \ref{fig:Fig2}(c). However, the experimental data in Fig \ref{fig:Fig2}(b) has data points spaced at 200ps. To mitigate errors in determining $\Omega(\mathrm{dt})$ at high rates, $\eta(\mathrm{dt})$ is interpolated and smoothed. Even so,  we find that Equation \ref{eqomega} displays numerical errors at rates $R > 10^9$ ($R' \sim 2 \times 10^7$). Therefore, we truncate our simulation at $R > 10^9$.

\section{Calculating the magnitude of the auto-correlation}
\label{sec:AppendixCorrelation}
It is well known that the second-order intensity correlation function is:
\begin{equation} 
g^{(2)}(\tau) = \dfrac{\langle I(t)I(t+\tau)\rangle}{\langle I(t)\rangle \langle I(t+\tau)\rangle},
\label{eq:g2}
\end{equation}
where the triangular brackets denote the mean value. By extension the $n$th order correlation function is
\begin{equation} 
g^{(n)}(\tau_1, \tau_2....\tau_{n-1}) = \dfrac{\langle I(t)I(t+\tau_1)...I(t+\tau_{n-1})\rangle}{\langle I(t)\rangle \langle I(t+\tau_1)\rangle ...\langle I(t+\tau_{n-1})\rangle}.
\label{eq:gN}
\end{equation}
However, in the experiment we correlate the detected light at rate $R'_{i}$ which is scaled down from the incident photon rate on the $i$th detector by the rate-dependent efficiency $\epsilon_{i}(R_{i})$ to give $R'_i = \epsilon_{i}(R_{i}) R_{i}$, assuming equivalence between $R$ and $I$. Therefore, the measured correlation function at coincidence is
\begin{equation} 
g^{(n)}_{\mathrm{exp}}(0,0...) = \dfrac{\langle  \prod^{n}_{i=1}\epsilon_{i}(R_{i})\cdot R_{i} \rangle}{ \prod^{n}_{i=1} \langle \epsilon_{i}(R_{i})\cdot R_{i} \rangle },
\label{eq:gN0}
\end{equation}
Which we simplify by assuming all detectors, $i$, have equal efficiencies $\epsilon_{i}(R)=\epsilon_{d}(R)$ and receive equal photon rates, $R$. 
\begin{equation} 
g^{(n)}_{\mathrm{exp}}(0,0...) = \dfrac{\langle  (\epsilon_{d}(R)\cdot R)^{n} \rangle}{ \langle \epsilon_{d}(R)\cdot R \rangle^{n} }
= \dfrac{\int^{\infty}_{0} [\epsilon_{d}(R)\cdot R]^{n}\cdot \xi(R)\cdot \mathrm{d}R}{[\int^{\infty}_{0} \epsilon_{d}(R)\cdot R\cdot \xi(R)\cdot \mathrm{d}R]^{n}}.
\label{eq:gN00}
\end{equation}

In Fig. \ref{fig:Fig3} we show the results of numerically evaluating Equation \ref{eq:gN00} using the previously determined detector efficiency factor, $\epsilon_{d}(R)$, and the well known intensity distribution of thermal light, $\xi(I) = \langle I \rangle^{-1} e^{\frac{-I}{\langle I\rangle}} $ and assuming $\xi(I)= \xi(R)$. Note that the data in Fig. \ref{fig:Fig3} is plotted as a function of the detected rate $R'$ (rather than the incident rate $R$).

Assuming an equal intensity split between identical detectors, and considering the case of low $R'$, we can make the approximation $\epsilon_{d}(R)=1$. Equation \ref{eq:gN} reduces to:
\begin{equation} 
g^{(n)}(0,0...) = \dfrac{\langle  I^{n} \rangle}{ \langle I \rangle^{n} },
\label{eq:7}
\end{equation}
which we can solve analytically using the intensity distribution of thermal light $P(I) = \langle I \rangle^{-1} e^{\frac{-I}{\langle I \rangle}} $. The numerator of Equation \ref{eq:7} is given by:
\begin{equation} 
\langle I^{n} \rangle = \dfrac{\int^{\infty}_{0} I^{n} \cdot \xi(I) \mathrm{d}I}{\int^{\infty}_{0} \xi(I) \mathrm{d}I} = n!\langle I \rangle^{n}.
\label{eq:8}
\end{equation}

Thus, $g^{(n)}(0,0...) = n!$ for thermal light in the regime where $\epsilon_{d}(R)=1$. However, when the detectors have a TER-related saturation the degree of correlation is intensity dependent, as we show in Fig. 3 of the manuscript. Our intuitive understanding of this effect can be explained with reference to Equation \ref{eq:gN0}: in the numerator high intensity bunches of photons are detected with reduced fidelity at higher flux, relative to denominator where the mean photon flux is lower. This reduces measured correlation function at high intensities and high levels of bunching. We note that for ideal antibunched sources, with $g^{(2)}(0) = 0$ the TER has no effect on the coincident values of the correlation because the source does not emit more than one photon at the same time but does result in a variable saturation curve for the anti-bunched source, as we show in Fig. \ref{fig:Fig1}. However, the effect can distort the measurement when $g^{(2)}(0)$ is non-zero but below 1 as discussed in Ann $et$ $al$ \cite{Ann2015}. We can also say that the Poissonian source will have a $g^{(2)}(0) = 1$ regardless of the TER because there is no intensity correlation, so the numerator and denominator of Equation \ref{eq:gN00} are equal.


%

\end{document}